\documentclass[aps,prb,twocolumn,superscriptaddress]{revtex4-1}

\usepackage{graphicx}
\usepackage{adjustbox}
\usepackage{chngcntr}
\usepackage{color}
\graphicspath{{~/data/article/}}

\begin{document}


\title{Canted antiferromagnetic order in the monoaxial chiral magnets V$_{1/3}$TaS$_{2}$ and V$_{1/3}$NbS$_{2}$}



\author{K. Lu}
\affiliation{Department of Physics and Materials Research Laboratory, University of Illinois at Urbana-Champaign, Urbana, IL 61801, USA}

\author{D. Sapkota}
\affiliation{Department of Physics and Astronomy, University of Tennessee, Knoxville, Tennessee 37996, USA}

\author{L. DeBeer-Schmitt}
\author{Y. Wu}
\author{H. B. Cao}
\affiliation{Neutron Scattering Division, Oak Ridge National Laboratory, Oak Ridge, Tennessee 37831, USA}

\author{N. Mannella}
\affiliation{Department of Physics and Astronomy, University of Tennessee, Knoxville, Tennessee 37996, USA}

\author{D. Mandrus}
\affiliation{Department of Materials Science and Engineering, University of Tennessee, Knoxville, TN 37996, USA}
\affiliation{Department of Physics and Astronomy, University of Tennessee, Knoxville, Tennessee 37996, USA}
\affiliation{Materials Science and Technology Division, Oak Ridge National Laboratory, Oak Ridge, TN 37831, USA}

\author{A. A. Aczel}
\email{aczelaa@ornl.gov}
\affiliation{Neutron Scattering Division, Oak Ridge National Laboratory, Oak Ridge, Tennessee 37831, USA}
\affiliation{Department of Physics and Astronomy, University of Tennessee, Knoxville, Tennessee 37996, USA}

\author{G. J. MacDougall}
\email{gmacdoug@illinois.edu}
\affiliation{Department of Physics and Materials Research Laboratory, University of Illinois at Urbana-Champaign, Urbana, IL 61801, USA}


\date{\today}

\begin{abstract}
The Dzyaloshinskii-Moriya (DM) interaction is present in the transition metal dichalcogenides (TMDC) magnets of form $M_{1/3}T$S$_{2}$ ($M$~$=$~3d transition metal, $T$~$\in$ \{Nb, Ta\}), given that the intercalants $M$ form $\sqrt{3}\times\sqrt{3}$ superlattices within the structure of the parent materials $T$S$_2$ and break the centrosymmetry. Competition between DM and ferromagnetic exchange interactions has been shown to stabilize a topological defect known as a chiral soliton in select intercalated TMDCs, initiating interest both in terms of fundamental physics and the potential for technological applications. In the current article, we report on our study of the materials V$_{1/3}$TaS$_2$ and V$_{1/3}$NbS$_2$, using a combination of x-ray powder diffraction, magnetization and single crystal neutron diffraction. Historically identified as ferromagnets, our diffraction results instead reveal that vanadium spins in these compounds are arranged into an A-type antiferromagnetic configuration at low temperatures. Refined moments are 1.37(6)$\mu_{B}$ and 1.50(9)$\mu_{B}$ for V$_{1/3}$TaS$_2$ and V$_{1/3}$NbS$_2$, respectively. Transition temperatures $T_{c}$~$=$~32K for V$_{1/3}$TaS$_{2}$ and 50K for V$_{1/3}$NbS$_{2}$ are obtained from the magnetization and neutron diffraction results. We attribute the small net magnetization observed in the low-temperature phases to a subtle ($\sim$2$^{\circ}$) canting of XY-spins in the out-of-plane direction. These new results are indicative of dominant antiferromagnetic exchange interactions between the vanadium moments in adjacent \textit{ab}-planes, likely eliminating the possibility of identifying stable chiral solitons in the current materials.
\end{abstract}


\maketitle

\section{Introduction}
\begin{figure}[h]
\includegraphics[width=0.5\textwidth]{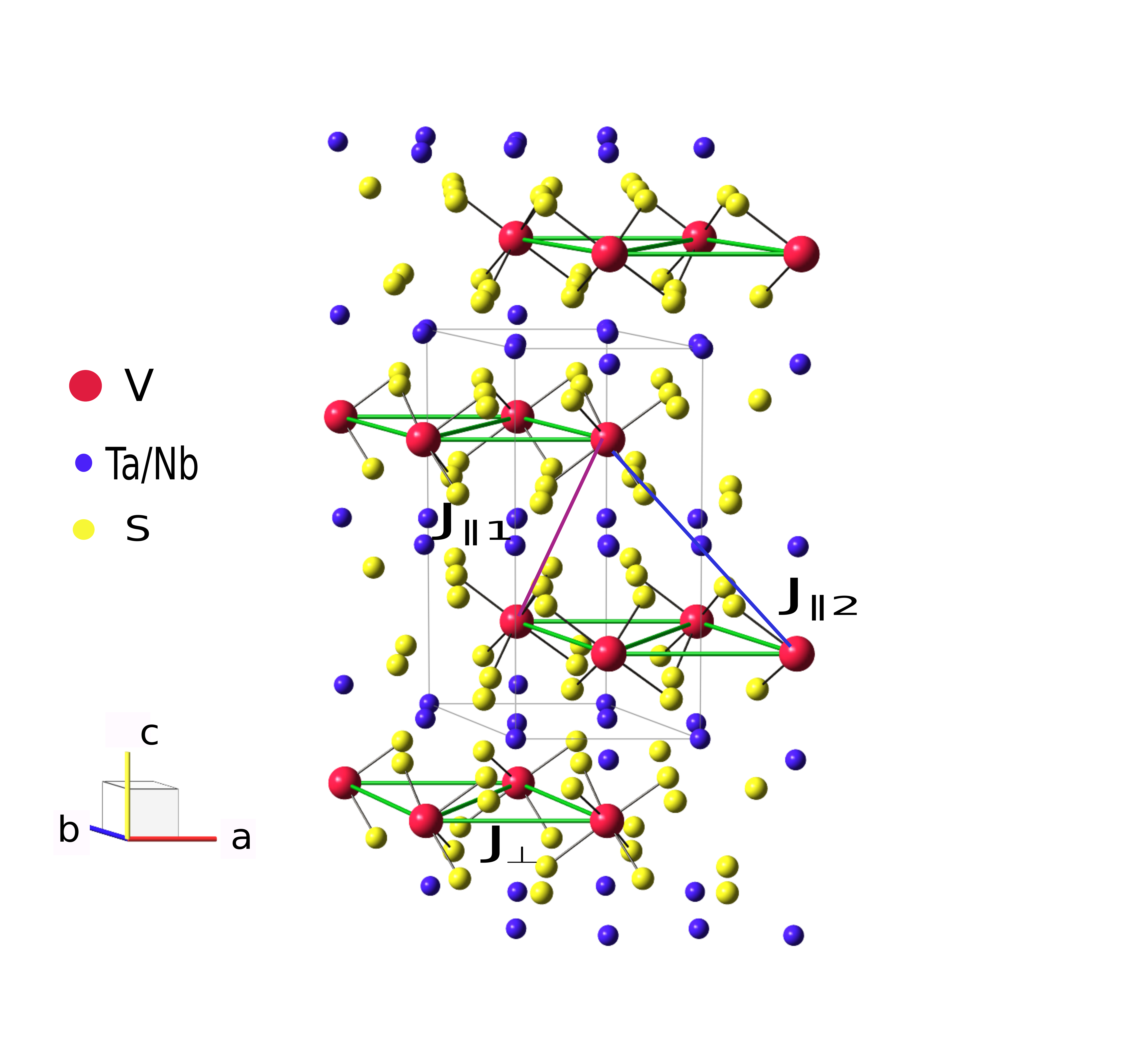}
 \caption{\label{fig:structure} Crystal structure of $M_{1/3}T$S$_2$ ($M$~$=$~3d transition metal; $T$~$\in$ \{Nb, Ta\}) with space group {\it P6$_{3}$22} and intercalants $M$ occupying 2c Wyckoff positions. The nearest-neighbor, next-nearest-neighbor, and third-nearest-neighbor exchange interactions (i.e. $J_\perp$, $J_{\parallel1}$, and $J_{\parallel2}$) between intercalants are indicated by green, magenta, and blue lines respectively. }
\end{figure}

The transition metal dichalcogenides (TMDC) have attracted a great amount of attention in recent years due to the unique physics demonstrated in several compounds and the promise of applicability in future technologies \cite{chhowalla_2013_the,manzeli_2017_2d}. Among these materials, intercalated variants $M_{x}T$S$_2$ ($M$~$=$~3d transition metal; $T$~$\in$ \{Nb, Ta\}) are of particular interest when $x$~$=$~1/3. In this case, the charged ion cores of the intercalated atoms order into a stacked  $\sqrt{3}\times\sqrt{3}$ superlattice, as shown in Fig.~\ref{fig:structure}, which lowers the crystal symmetry \cite{parkin_1980_3d,friend_1977_electrical,hulliger_1970_on,parkin_1980_magnetic,vandenberg_1968_structural,vanlaar_1971_magnetic}. The intercalated atoms are typically either trivalent or divalent \cite{parkin_1980_3d}, and the orbital moment is usually quenched due to the octahedral crystal fields from neighboring sulfur atoms. Therefore, the localized magnetic moment at the intercalant site originates mainly from the spins of the unpaired electrons \cite{friend_1977_electrical,mito_2019_observation,polesya_2019_electronic}. The host materials TaS$_2$ and NbS$_2$ are layered TMDCs of 2H polytype and provide conduction electrons to the entire system. Previous work has shown that the exchange interactions in this isostructural family are quite long-ranged and variable in both sign and magnitude \cite{aczel_2018_extended, 16_mankovsky}, which support a dominant RKKY mechanism and allow for a rich and diverse collection of magnetic ground states. Extensive investigation of these materials in the 1970s, mostly with magnetization, led to reports that the materials with $M$~$\in$ \{V, Cr, Mn\} compounds are ferromagnetic, that materials with $M$~$\in$ \{Co, Ni\} compounds are antiferromagnetic, and that materials with $M$~$=$~Fe can be either depending on the host TMDC \cite{parkin_1980_3d}. More recently, intriguing chiral magnetism observed in some materials \cite{togawa_2012_chiral,braam_2015_magnetic,kousaka_2016_long,kousaka_2009_chiral,dai_2019_critical,karna_2019_consequences} and discussion of future technological applications \cite{borisov_2009_magnetic,kishine_2011_tuning,okumura_2019_spincurrent,nair_2019_electrical} have resulted in an increasing demand for detailed microscopic information about the magnetic order across the entire $M_{1/3}$TS$_2$ family.



The onset of chiral magnetism is rooted in the non-centrosymmetric structure of the $M_{1/3}$TS$_2$ materials. In ideal samples, these materials have space group {\it P6$_3$22} and intercalants occupying 2c Wyckoff positions, which modifies the placement of neighboring sulfur atoms in a way that induces a chiral axis along the c-direction, due to the lack of inversion symmetry and a mirror plane. This chirality allows a non-negligible contribution of the antisymmetric exchange, also known as the Dzyaloshinskii-Moriya (DM) interaction, to the magnetic Hamiltonian \cite{moriya_1960_anisotropic,moriya_1982_evidence}. Theoretically, a combination of Heisenberg exchange and the DM interaction can give rise to modulated magnetic structures such as conical, helical and skyrmion states \cite{rler_2006_spontaneous,seki_2012_observation,muhlbauer_2009_skyrmion,rybakov_2015_new}. Skyrmions are often treated as emergent particles and are topologically stable \cite{rler_2006_spontaneous}, which suggests great advantages in future applications in spintronics \cite{zhang_2015_magnetic,woo_2018_deterministic}. For this reason, the past decade has seen an intense search for new skyrmion materials with variable properties, though most candidates to-date are cubic chiral magnets \cite{muhlbauer_2009_skyrmion,yu_2010_realspace}.

In intercalated TMDCs, the layered structure results in strong magnetic anisotropy along the out-of-plane direction, leading these materials to be known as monoaxial chiral magnets, distinct from the cubic systems. The full details of the magnetic anisotropy are difficult to account for and often depend on both the host material and the intercalants. In the case where the out-of-plane direction is the hard axis (i.e. easy-plane anisotropy), the ordered spin configuration can manifest as a chiral helical magnetic state (CHM) if the exchange interaction $J$ along the out-of-plane direction is ferromagnetic \cite{togawa_2012_chiral,togawa_2016_symmetry}. The exchange interactions tend to ferromagnetically align the neighboring moments while the DM interaction $\bf{D}$ favors perpendicular alignment of them between planes \cite{Ijiri_2007_link}. Therefore, a helical ground state with the helical axis along the c-direction and the spins rotating in the \textit{ab}-plane is expected. The helical pitch is set by the relative strength between $J$ and $\bf{D}$ \cite{togawa_2012_chiral}, assuming that there is only one out-of-plane exchange constant. Previous experimental \cite{aczel_2018_extended} and theoretical works \cite{16_mankovsky} have shown that the true situation in intercalated TMDCs is somewhat more complicated, as exchange interactions up to at least third nearest neighbor are important. These interactions are shown in Fig.~\ref{fig:structure} and consist of one in-plane and two out-of-plane contributions, labeled as $J_\perp$, $J_{\parallel1}$, and $J_{\parallel2}$ to differentiate between terms connecting atoms with ($\parallel$) or without ($\perp$) a position vector component along the helical c-axis. Notably, all three interactions are FM in Cr$_{1/3}$NbS$_{2}$ and Mn$_{1/3}$NbS$_{2}$ and therefore a competition between $J_{\parallel1}$, $J_{\parallel2}$, and $\bf{D}$ drives the helical order.

In intercalated TMDCs, the helical state was proposed theoretically in 1982 \cite{moriya_1982_evidence} and quickly observed experimentally with the small angle neutron scattering (SANS) technique in Cr$_{1/3}$NbS$_{2}$ \cite{miyadai_1983_magnetic}. This state was further confirmed with Lorentz microscopy, small angle electron scattering and $\mu$SR on the same compound \cite{togawa_2012_chiral,braam_2015_magnetic}. Upon application of field perpendicular to the helical axis, theory predicts that ferromagnetic domains are enlarged and partitioned with evenly distributed 2$\pi$ phase twists, a phase known as a ``chiral soliton lattice'' (CSL) state \cite{togawa_2016_symmetry,kishine_2015_theory}. These 2$\pi$ phase twists (solitons) are emergent, topologically non-trivial particles similar to skyrmions, but only one dimensional (1D) in nature. Increasing the perpendicular field is predicted to elongate ferromagnetic domains and hence increase the spatial periodicity of the CSL. In other words, the CSL is an emergent, mesoscopic crystalline system with a lattice constant which is tunable with an applied field. Evidence for the CSL state has been seen experimentally in bulk magnetization and transport measurements, but to-date convincingly identified only in the singular compound Cr$_{1/3}$NbS$_{2}$ \cite{ghimire_2013_magnetic,clements_2017_critical}. Thus, Cr$_{1/3}$NbS$_{2}$ is regarded as a prototype for studying soliton physics, and a plethora of studies have been devoted to this compound to characterize its topological robustness and tunability with field, current and pressure \cite{clements_2017_critical,han_2017_tricritical,masaki_2018_chiral,togawa_2013_interlayer,mito_2015_investigation}. The relative stability of skyrmion and soliton lattices is believed to be strongly associated with the magnetic anisotropy \cite{butenko_2010_stabilization,karhu_2012_chiral,wilson_2014_chiral}, and a systematic study across all previously mentioned intercalated TMDCs to elucidate this relationship is desirable.

Recent reports along this theme include a single neutron diffraction study of polycrystalline Cr$_{1/3}$TaS$_{2}$, which suggested this material may be a second soliton candidate \cite{kousaka_2016_long}, and multiple magnetization and diffraction studies of Mn$_{1/3}$NbS$_{2}$, which show evidence of long period spin modulation \cite{kousaka_2009_chiral,dai_2019_critical,karna_2019_consequences}. These findings firmly establish the ferromagnetic TMDC compounds as strong candidates for hosting solitons and motivate further work. In the following article, we present magnetization and neutron diffraction results on single crystals of the vanadium-intercalated compounds V$_{1/3}$TaS$_{2}$ and V$_{1/3}$NbS$_{2}$, two other materials historically identified as ferromagnets \cite{parkin_1980_3d}. In contrast to this designation, we show that our data instead implies A-type antiferromagnetic order in these systems, composed of a staggered stacking of ferromagnetic planes. A small net moment is observed with magnetization, consistent with previous results \cite{nakanishi_2008_structural}, which we propose is associated with a small out-of-plane canting of the XY-like spins. As part of a more general discussion, we will also address some common points of confusion regarding the crystal structure of these compounds, and lay out our methodology to discriminate between well-ordered samples with non-centrosymmetric {\it P6$_3$22} symmetry and inversion-preserving disordered samples, which have distinctly different magnetic properties.

\section{Experimental Methods}

The vanadium intercalated TMDC compounds explored in this study were grown via a two-step process. In the first step, polycrystalline samples were obtained by sintering a stoichiometric mixture of V (Sigma-Aldrich, 99.5\% trace metal basis),  Nb (Alfa Aesar, 99.8\% trace metal basis)/Ta (Sigma-Aldrich, 99.99\% trace metal basis) and S (Alfa Aesar, 99.99\% metal basis) powders in evacuated quartz ampoules at 900$^{\circ}$C for a week. Single crystals were grown from these polycrystals via a chemical vapor transport method using iodine as the transport agent in evacuated quartz tubes. The quartz tubes were subjected to a temperature gradient from 950$^{\circ}$C to 800$^{\circ}$C for one week to allow formation of single crystals with an average lateral dimension $\sim$3mm. The chemical composition of the single crystals was confirmed by the energy dispersive x-ray technique.

Both polycrystalline samples and single crystals were used for magnetization and x-ray measurements as discussed in this article. Single-crystal x-ray diffraction measurements were performed on crystalline samples of V$_{1/3}$TaS$_{2}$ using a Bruker X8ApexII four-circle kappa-axis diffractometer at the George L. Clark X-Ray Facility in Noyes Laboratory at Illinois. An incident x-ray beam with wavelength 0.7107~\AA~was selected using a graphite monochromator. The measured samples had dimensions $\sim$ 200$\mu$m$\times$200$\mu$m. Reflections were collected at room temperature over a 2$\theta$ range of 12.4$^{\circ}$ to 72.6$^{\circ}$ with a resolution of $\sim$ 0.6 \AA. Diffraction patterns were refined to determine lattice symmetry. We followed our single crystal measurements with the acquisition of x-ray powder diffraction patterns from a small volume of polycrystalline samples of both materials. Powder measurements were performed using a Bruker D8 Advance x-ray Diffractometer with a 1.5406~\AA~wavelength incident beam at room temperature. Magnetization measurements were performed on a number of different powder samples as a function of temperature in an applied field $H$~$=$~500 Oe using a Quantum Design MPMS3 SQUID magnetometer housed in the Illinois Materials Research Laboratory.

Single crystals were also probed by neutron diffraction to determine the magnetic structure. Single crystal neutron diffraction data were collected using the HB-3A four-circle diffractometer at the High Flux Isotope Reactor in Oak Ridge National Laboratory. The samples were glued on an aluminum holder and mounted in a closed-cycle helium-4 refrigerator with a temperature range from 4 to 450K with no external magnetic field applied. A bent silicon monochromator selected out a monochromatic incident beam of wavelength 1.546~\AA. Scattering intensities at several Bragg peak locations were measured using an area detector at temperatures of 4.5K and 50K for V$_{1/3}$TaS$_2$ and at 4.5K and 70K for V$_{1/3}$NbS$_2$.

\section{Magnetization and X-ray diffraction}

As discussed in the preceding sections, the origin of chiral magnetism and the stabilization of CSL ground states depend critically on the formation of non-centrosymmetric crystal structures, which in the current materials are the result of intercalants ordering onto the 2c Wyckoff site of the space group {\it P6$_{3}$22}. Disorder among intercalants is expected to suppresses the DM interaction, and recent work on Cr$_{1/3}$NbS$_{2}$ correlates site disorder with not only a different crystal structure, but a purely ferromagnetic (rather than chiral) spin configuration with a significantly lower transition temperature \cite{dyadkin_2015_structural}. With these results in mind, our investigation of V$_{1/3}$TaS$_{2}$ and V$_{1/3}$NbS$_{2}$ began with a combined x-ray diffraction and magnetization study of the materials' properties to confirm the non-centrosymmetric space group and expected magnetic transition temperatures.

As a first step, single crystal x-ray diffraction was performed, as laid out above. Diffraction patterns were refined to determine symmetry. Results, however, were complicated by the existence of structural domains, and fitted equally well to several different space groups including {\it P6$_{3}$22} and {\it P6$_{3}/$mmc}. Lattice and goodness-of-fit parameters are shown for each of these refinements in Table.~\ref{table:x-ray}. As such domain formation results naturally as part of the crystal growth process, we determined this ambiguity likely represented a fundamental limitation of single crystal diffraction, making it unsuitable for determining centrosymmetry.

\begin{table}
\caption{\label{table:x-ray} Refined lattice parameters from x-ray diffraction data of powders of both materials and single crystals of V$_{1/3}$TaS$_{2}$ using both the non-centrosymmetric space group {\it P6$_{3}$22} and the centrosymmetric space group {\it P6$_{3}/$mmc}. All lattice constants are given in units of \AA. R-factors for each structural model are shown.}

\begin{ruledtabular}
\begin{tabular}{|c|c|c|c|}
\hline
 Model & V$_{1/3}$NbS$_{2}$ & V$_{1/3}$TaS$_{2}$ & V$_{1/3}$TaS$_{2}$ \\
		  & Powder & Powder & Crystal \\
\hline
{\it P6$_3$22} & a = 5.7564(3) & a = 5.72960(7) & a = 5.7196(2)\\
               & c = 12.151(1) & c = 12.1829(3) & c = 12.1665(4)\\
               & (R = 7.59\%)  & (R = 8.09\%)  & (R = 1.38\%) \\
\hline
 {\it P6$_3$/mmc} & a = 3.3232(1) & a = 3.30813(4) & a = 3.3022(1)\\
                  & c = 12.1479(9) & c = 12.1829(3) & c = 12.1665(4) \\
                  & (R = 8.13\%)  & (R = 7.57\%)  & (R = 1.25\%) \\
\hline
\end{tabular}
\end{ruledtabular}
\end{table}

\begin{figure*}[htbp]
\includegraphics[width=1\textwidth]{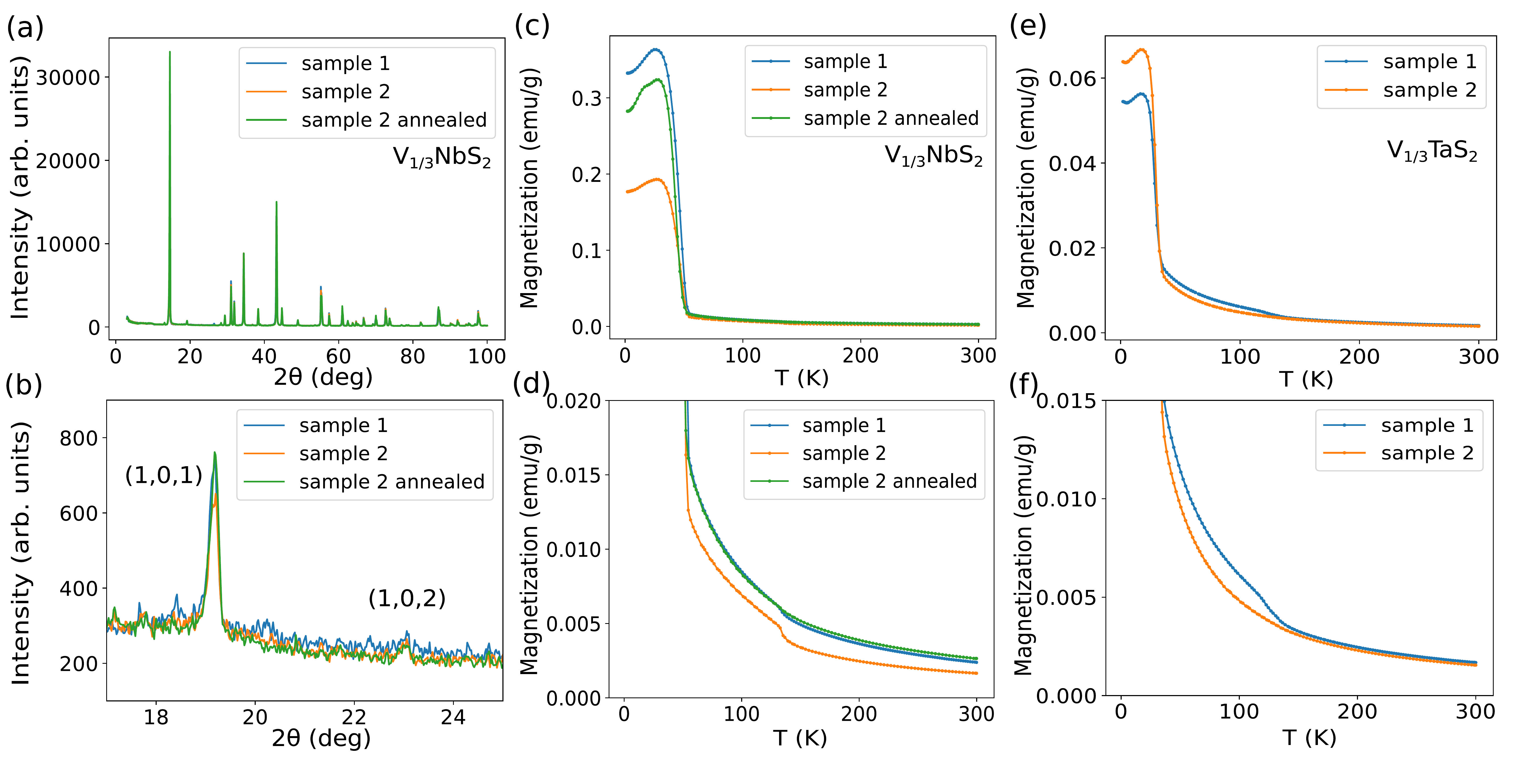}
 \caption{\label{fig:magnetization}(a) Powder x-ray diffraction pattern of V$_{1/3}$NbS$_{2}$. (b) The variation of the (1,0,1) structural Bragg peak intensity for different samples of V$_{1/3}$NbS$_{2}$. Sample 2 shows a decreased intensity for the (1,0,1) Bragg peak, as well as a weak high temperature magnetic transition $\sim$ 140K. (c), (e) Comparison of powder magnetization data for different samples of V$_{1/3}$NbS$_{2}$ and V$_{1/3}$TaS$_{2}$, respectively. (d), (f) Zoomed-in views of the transition at $\sim$ 140K that is only visible for sample 2 of V$_{1/3}$NbS$_{2}$ and sample 1 of V$_{1/3}$TaS$_{2}$. This high-temperature transition in V$_{1/3}$NbS$_{2}$ is suppressed after annealing, as shown by the green curve in (d).}
\end{figure*}

We thus followed our single crystal measurements with the acquisition of x-ray powder diffraction patterns from a small volume of polycrystalline samples. Resultant patterns for V$_{1/3}$NbS$_2$ are shown in Fig.~\ref{fig:magnetization}(a), and similar to the single crystal results, were equally well described by the non-centrosymmetric space group {\it P6$_{3}$22} and the centrosymmetric {\it P6$_{3}/$mmc} space group associated with the underlying transition metal dichalcogenide (TMDC) materials. The refined lattice parameters are shown in Table~\ref{table:x-ray}, and are in agreement with both the single crystal x-ray results and previous literature \cite{parkin_1980_3d}. The comparable success of both fitting models is due to the fact that inversion symmetry breaking in these compounds is entirely due to the relatively light intercalant atoms, and therefore the pertinent scattering features allowing differentiation between the various space groups are relatively weak. Nonetheless, we managed to identify some of the unique scattering features associated with the {\it P6$_{3}$22} space group in our powder diffraction patterns and correlate them with the bulk magnetic properties at low temperature, as described in more detail below.

The main results of magnetization measurements are shown in Fig.~\ref{fig:magnetization}(c) and (e). All samples investigated showed evidence of a magnetic transition, specifically at 32K for V$_{1/3}$TaS$_{2}$ and 50K for V$_{1/3}$NbS$_{2}$, marking the onset of a weak net moment of  0.01~$\mu_B$ and 0.03~$\mu_B$ per vanadium spin, respectively. These numbers are much smaller than the expected spin-only moment size of 2~$\mu_B$ for V$^{3+}$ and consistent with a canted antiferromagnetic state, as we discuss below. In a small subset of powder samples, we also observed a second transition at the higher temperature $T_{c}'$~$>$~100K, as shown in Fig.~\ref{fig:magnetization}(d) and (f). The co-existence of two ordering temperatures draws immediate parallels to the case of Cr$_{1/3}$NbS$_{2}$, where it has been shown that the configuration of the intercalants (i.e. ordered vs disordered) plays a major role in establishing the ordering temperature \cite{dyadkin_2015_structural}.

In support of this idea, we further observed an inverse correlation between the strength of the susceptibility signature for the high-temperature magnetic transition in V$_{1/3}$NbS$_{2}$ and the intensity of the (1,0,1) structural Bragg peak, which we explain in the Appendix is a signature of broken inversion symmetry. This is seen in Fig.~\ref{fig:magnetization}(b), where a decrease in the (1,0,1) scattering intensity is seen in the sample containing a second higher temperature transition (sample 2). Furthermore, when the same sample was annealed at a temperature of 900$^{\circ}$C for 24hrs, allowing the intercalants to order, both the higher temperature transition is suppressed and the intensity of the (1,0,1) Bragg peak is increased. A similar double transition is seen in a subset of V$_{1/3}$TaS$_{2}$ powders, though due to increased statistical uncertainty, we were unable to definitively confirm the same variation in the (1,0,1) Bragg peak intensity in this material. This is due to both an increase in scattering background associated with heavier Ta cations and a decrease in the volume associated with the high temperature transition. Nevertheless, given these collective results, we believe that the lowest temperature transition observed with magnetization is correlated with the desirable intercalant ordering in these compounds, and can be used to confirm the correct non-centrosymmetric space group.

\section{Zero-Field Single Crystal Neutron Diffraction}

We further determined the microscopic nature of the low-temperature ordered states based on the single-crystal neutron diffraction data collected. Rocking curve scans around each Bragg peak were first merged and then fitted to a Gaussian profile (Gaussian fitting method) using the HFIR Single Crystal Reduction Interface of Mantid \cite{arnold_2014_mantiddata} to extract the integrated intensities for both samples. For V$_{1/3}$TaS$_2$, the integrated intensities were also obtained using a simple summation method. Both methods (Gaussian fitting and simple summation) produce comparable results and lead to the same conclusions. The refined results based on the simple summation method are shown in this article for V$_{1/3}$TaS$_2$. Once the integrated intensities were obtained, the software then automatically corrected the intensities by dividing by the appropriate Lorentz factors for each Bragg peak to obtain the structure factor $F^{2}(\bf{\vec{Q}})$. Refinements of the extracted structure factors were performed using the FullProf software suite \cite{FULLPROF}.

\begin{figure*}[htbp]
 \centering
 \includegraphics[width=1\textwidth]{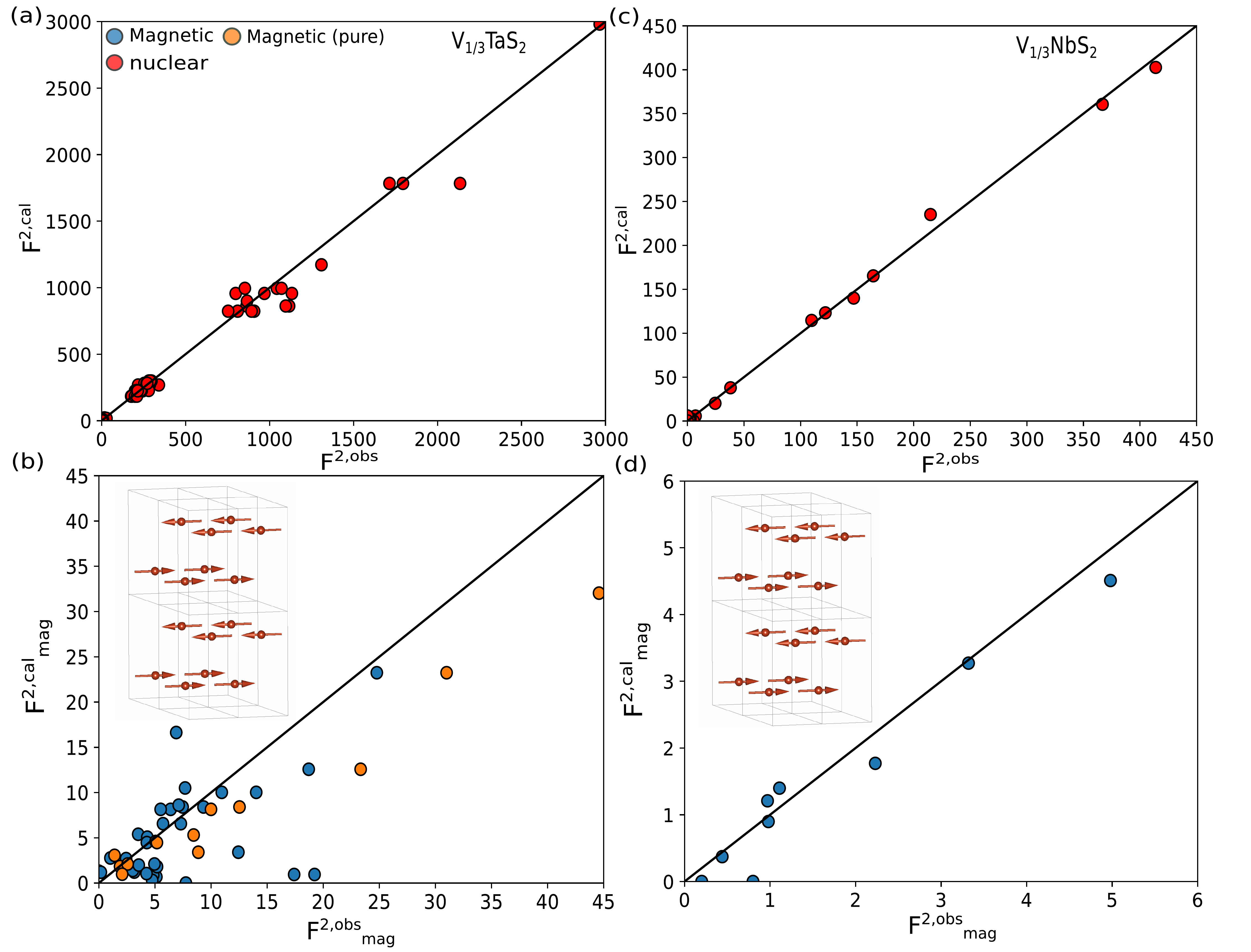}
 \caption{\label{fig:fit} (a), (c) Calculated nuclear structure factors $F^{2, cal}$ based on the non-centrosymmetric {\it P6$_{3}$22} model vs the observed nuclear structure factors $F^{2, obs}$ for V$_{1/3}$TaS$_{2}$ and V$_{1/3}$NbS$_{2}$, respectively. A 45$^{\circ}$ line is drawn as a reference. (b), (d) Calculated magnetic structure factors $F^{2, cal}_{mag}$ based on the $\Gamma_{6}$ model vs the observed magnetic structure factors $F^{2, obs}_{mag}$ for both materials. The orange data points in the V$_{1/3}$TaS$_{2}$ plot correspond to Bragg peaks with no intensity above the magnetic transition temperature (i.e. purely magnetic).}
 \end{figure*}

As the first step of the refinement, the high temperature data were used to check sample quality and confirm the expected crystal structure. In particular, these data are well-described by space group {\it P6$_3$22} reported previously (with R-factors being 5.93\% and 2.47\% for V$_{1/3}$TaS$_{2}$ and V$_{1/3}$NbS$_{2}$), as shown in Fig.~\ref{fig:fit}(a) and (c), respectively. As the element vanadium is a predominantly incoherent scatterer of neutrons, nothing further can be said about the positions of the intercalants within this structure. The vanadium is assumed to be at the 2c Wyckoff positions, however, based on the x-ray and magnetization data described above and the relevant literature \cite{parkin_1980_3d,friend_1977_electrical,hulliger_1970_on,parkin_1980_magnetic,vandenberg_1968_structural,vanlaar_1971_magnetic}.

\begin{figure*}[htbp]
 \includegraphics[width=1\textwidth]{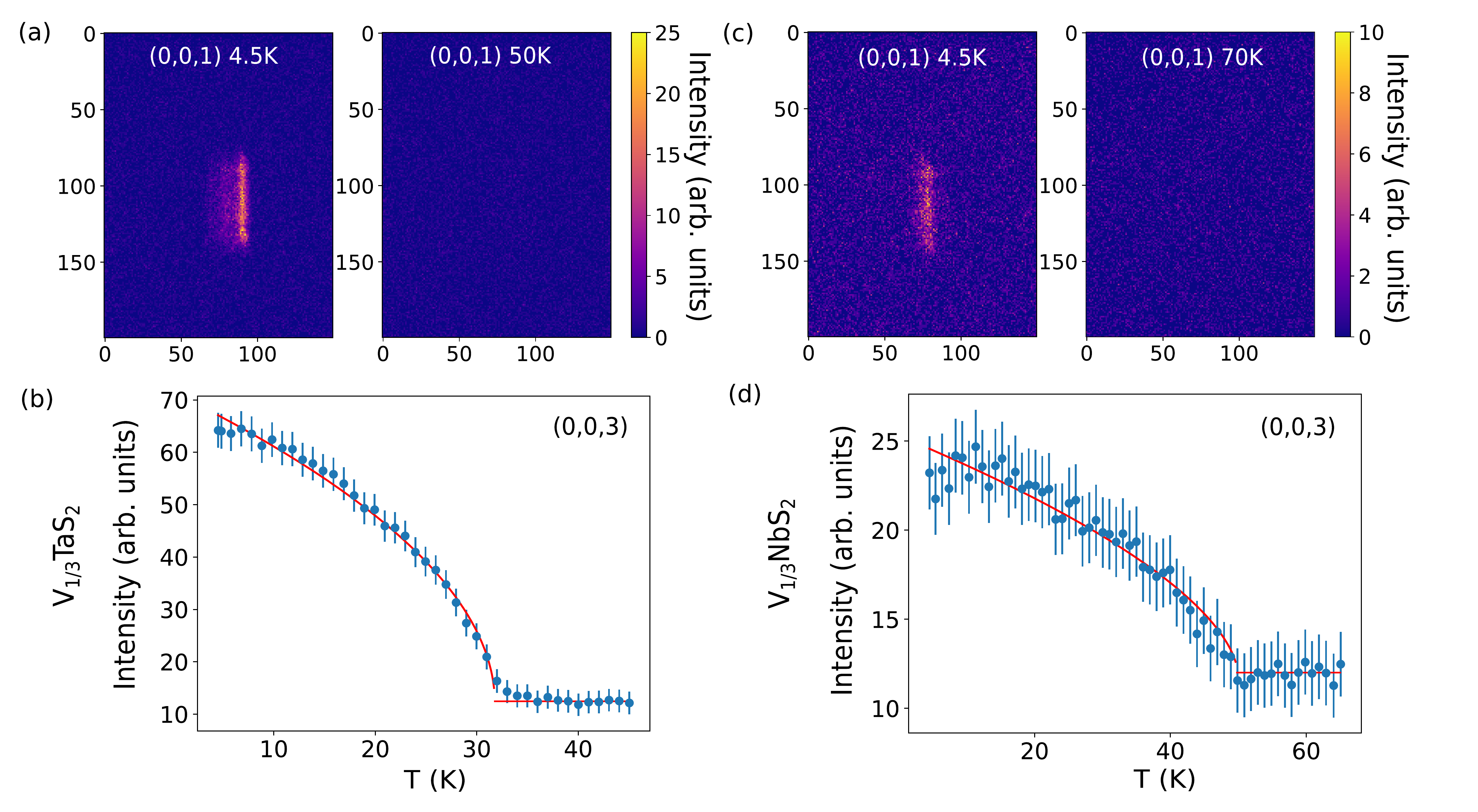}
 \caption{\label{fig:OP} (a), (c) Area detector images of the (0,0,1) Bragg peak for V$_{1/3}$TaS$_{2}$ and V$_{1/3}$NbS$_{2}$ below and above the magnetic transition, respectively. (b), (d) Order parameter scan of the (0,0,3) Bragg peak for V$_{1/3}$TaS$_{2}$ and V$_{1/3}$NbS$_{2}$, with the power law fits described in the text superimposed on the data.}
 \end{figure*}

At 4.5K, excess scattering is observed in both materials at positions commensurate with nuclear Bragg peaks, as shown in Fig.~\ref{fig:OP}(a) and (c). Based on the symmetry of the space group {\it P6$_3$22}, the predicted nuclear structure factors of certain Bragg peaks, such as (0,0,1) and (0,0,3), are zero at high temperature. The observation of clear peaks at these positions for $T$~$=$~4.5K can therefore be uniquely associated with magnetic order, and implies that the magnetic structure is described by the propagation vector $\bf{\vec{k}}=$(0,0,0). Given that there are two vanadium moments per chemical unit cell, this propagation vector can in principle be associated with either a ferromagnetic or an antiferromagnetic ordering configuration.

In Fig.~\ref{fig:OP}(b) and (d), we show the temperature-dependent intensities of the (0,0,3) Bragg peak for both materials, which can be treated as an order parameter for the magnetic state. Fitting these curves to the functional form $ I(\bf{\vec{Q}})$ $= I_{0} + A \mid \frac{T-T_{c}}{T_{c}} \mid^{\alpha} $, we extracted numbers $T_c$ = 31.77(7) K and 49.6(5) K for the transition temperature of V$_{1/3}$TaS$_{2}$ and V$_{1/3}$NbS$_{2}$, respectively. These values show good agreement with the transitions inferred from our low-temperature susceptibility data.

\begin{figure}[htbp]
 \includegraphics[width=0.3\textwidth]{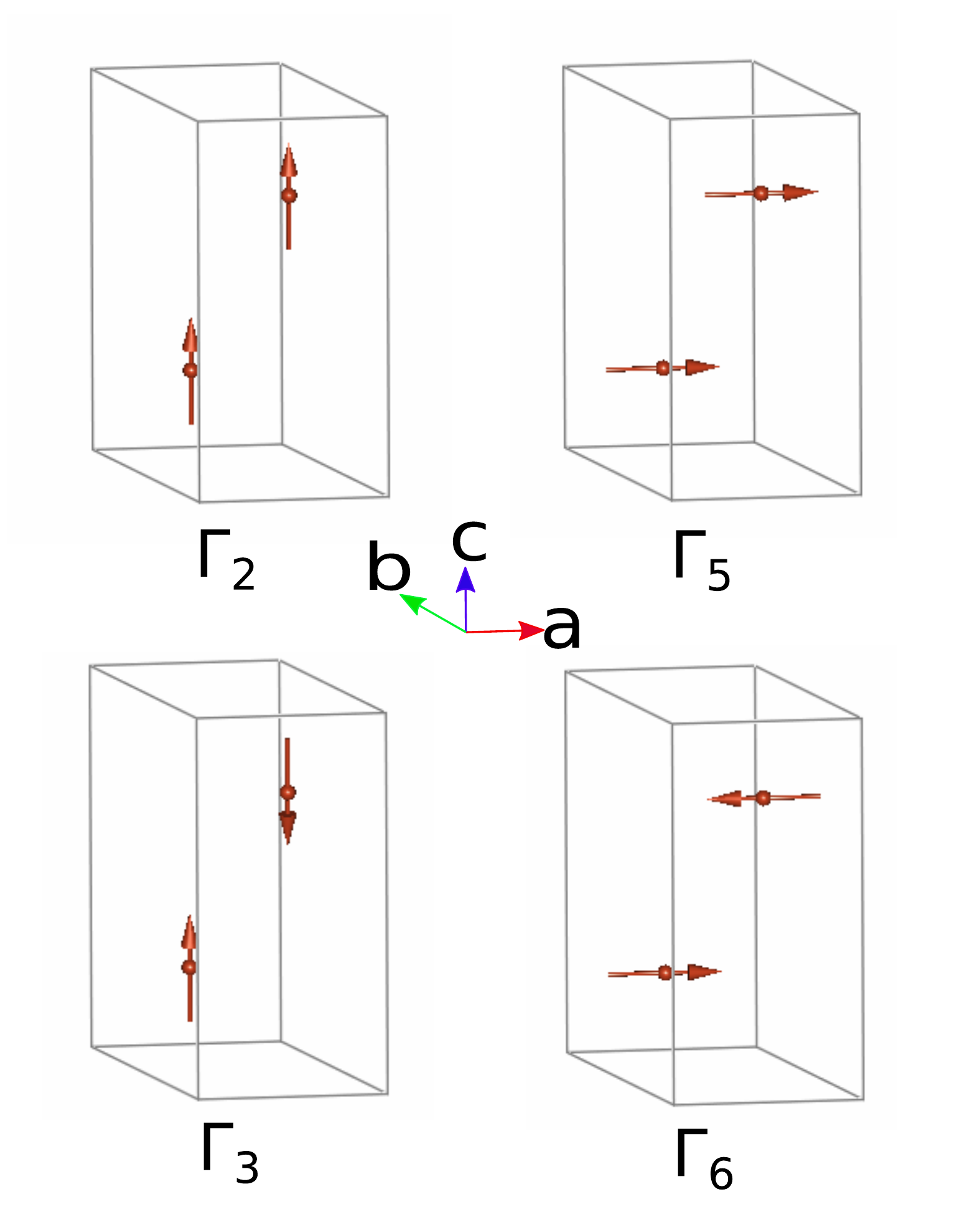}
 \caption{\label{fig:representation}Schematics depicting the ordered spin configurations for four different IRs of space group {\it P6$_{3}$22}, $\vec{k}=$(0,0,0) and vanadium at Wyckoff site 2c. The $\Gamma_{2}$ and $\Gamma_{3}$ representations consists of basis vectors along the c-axis, while the ordered moment is confined to the ab-plane for $\Gamma_{5}$ and $\Gamma_{6}$. }
 \end{figure}

To facilitate refinement of the magnetic structures, we extracted structure factors for both high and low temperature data, denoted $F^{2}_{HT}$ and $F^{2}_{LT}$, via the HFIR Single Crystal Reduction Interface of Mantid \cite{arnold_2014_mantiddata}. For each reciprocal lattice vector $\bf{\vec{Q}}$~$=$~(H,K,L) (r.l.u.) measured, the magnetic structure factor was defined as $F^{2}_{mag}(\bf{\vec{Q}})$ $=F^{2}_{LT}(\bf{\vec{Q}})$ $-F^{2}_{HT}(\bf{\vec{Q}})$, and associated uncertainties were obtained by adding the errors of the high and low temperature contributions in quadrature. A list of magnetic structure factors are shown in Table~\ref{table:structure_factor}. Candidate magnetic models were then obtained with the group representational analysis program SARAh \cite{SARAh}. For space group {\it P6$_{3}$22}, $\bf{\vec{k}}=$(0,0,0) and vanadium at Wyckoff site 2c, there are four irreducible representations (IRs): $\Gamma_{2}$, $\Gamma_{3}$, $\Gamma_{5}$ and $\Gamma_{6}$. Schematics of the spin configurations obtained from these four IRs are shown in Fig.~\ref{fig:representation}. We find that the pair $\{\Gamma_{2}, \Gamma_{3}\}$  ($\{\Gamma_{5}, \Gamma_{6}\}$) describes structures with moments along the c-axis (within the ab-plane), while the first IR corresponds to ferromagnetic alignment of spins in adjacent planes and the second IR corresponds to an antiferromagnetic stacking of ferromagnetic planes.

\begin{table}
\caption{\label{table:structure_factor} The observed and calculated magnetic structure factors for both V$_{1/3}$NbS$_2$ and V$_{1/3}$TaS$_2$. The calculated values are based on the $\Gamma_{6}$ magnetic structure model.}
\begin{ruledtabular}
\begin{tabular}{|c|c|c|c|c|}
\hline
 Bragg peak & $F^{2, obs}_{mag}$ & $F^{2, cal}_{mag}$ & $F^{2, obs}_{mag}$ & $F^{2, cal}_{mag}$ \\
                  & V$_{1/3}$NbS$_2$   & V$_{1/3}$NbS$_2$   & V$_{1/3}$TaS$_2$   & V$_{1/3}$TaS$_2$ \\
\hline
 (0,0,1) & 4.98(14) & 4.3768 & 44.59(337) & 44.4715 \\
 (0,0,3) & 3.32(7)   & 3.1755 & 31.00(292) & 32.2676 \\
 (0,0,5) & 2.23(9) & 1.7182 & 23.34(410) & 17.4602 \\
 (0,0,7) & 0.51(104) & 0.7257 & 8.45(460) & 7.3742 \\
 (1,0,0) & 0.98(3) & 0.8630 & 8.87(212) & 8.5394 \\
 (1,0,1) & 0.44(137) & 0.3585 & 1.89(212) & 3.5831 \\
 (1,0,2) & 1.11(4) & 1.3528 & 12.54(269) & 13.6266 \\
 (1,0,3) & 0.17(2) & 0.4556 & 1.40(291) & 4.6109  \\
 (1,0,4) & 0.97(5) & 1.1718 & 10.00(363) & 11.8717 \\
 (1,0,5) & 0.55(5)  & 0.2980 & 2.57(354) & 3.0241 \\
 (1,0,6) & 0.00(213)  & 0.6266  & 5.18(401) & 6.3575 \\
\hline
\end{tabular}
\end{ruledtabular}
\end{table}

We performed refinements with all four models using the FullProf software suite, and fits which used solely $\Gamma_{2}$, $\Gamma_{3}$ or $\Gamma_{5}$ IRs were of significantly poorer quality ($R>65\%$) than the fit using $\Gamma_{6}$ ($R=33.7\%$ for V$_{1/3}$TaS$_{2}$ and $R=6.8\%$ for V$_{1/3}$NbS$_{2}$). In Fig.~\ref{fig:fit}(b) and (d), we plot the calculated structure factor $F^{2, cal}_{mag}$ for the best fits based on the $\Gamma_{6}$ model against the observed structure factor $F^{2, obs}_{mag}$ described above for V$_{1/3}$TaS$_{2}$ and V$_{1/3}$NbS$_{2}$, respectively. For the refinement of V$_{1/3}$TaS$_{2}$, we noticed that there were many magnetic structure factors $F^{2, obs}_{mag}$ carrying large errors $\delta F^{2,obs}_{mag}$. However, these large errors are simply due to the inclusion of a number of magnetic peaks which are coincident with intense nuclear Bragg peaks. The resultant large statistical error bars for $\delta F^{2, obs}_{mag}$ decrease the precision of the magnetic scattering intensity, and as a result are largely excluded from the refinement in the FullProf fits. Restricting our attention to the purely magnetic peaks listed in Table~\ref{table:structure_factor} (orange points in Fig.~\ref{fig:fit}(b)), we obtain a much improved R-factor of 4.72$\%$ for the magnetic refinement of V$_{1/3}$TaS$_{2}$ with the $\Gamma_{6}$ model, and notably this model still provides the best description of the data. The refined vanadium magnetic moment is 1.37(6) $\mu_{B}$ for V$_{1/3}$TaS$_{2}$ and 1.50(9) $\mu_{B}$ for V$_{1/3}$NbS$_{2}$.

These refinements indicate that the ordered state of both V$_{1/3}$TaS$_{2}$ and V$_{1/3}$NbS$_{2}$ is best described as an antiferromagnetic stacking of ferromagnetic planes of spins, which implies a net moment of zero. To account for the net moment observed in the magnetization measurement, we considered a linear combination of $\Gamma_{6}$ with one of the two out-of-plane IRs, $\Gamma_{2}$ or $\Gamma_{3}$, corresponding to the ferromagnetic or antiferromagnetic canting of spins out of the plane, respectively. (We note that antiferromagnetic canting ideally still results in zero net moment, however the possibility of stacking faults cannot be ruled out, which can induce a small net moment.) The R-factors for fitting with either $\Gamma_{2}+\Gamma_{6}$ or $\Gamma_{3}+\Gamma_{6}$ were not significantly improved over fitting with a purely $\Gamma_{6}$ model, and refinements with either model indicate a canting angle $\sim$10$^{\circ}$, which would produce a magnetization of 1.8 $emu/g$ that is much too large to be reconciled with our magnetization measurements. For this reason, we conclude that the current neutron diffraction data set is not able to unambiguously determine the level of spin canting in these materials. It is worth noting that the net moment observed with bulk magnetization measurements is extremely small, which amounts to a moment of $\le$ 0.05 $\mu_{B}$ per vanadium, equivalent to a canting angle of $\le$ 2$^{\circ}$.

Previous experimental \cite{aczel_2018_extended} and theoretical work \cite{16_mankovsky} has shown that the exchange interactions in the intercalated TMDC family are highly-tunable and significant up to at least third nearest neighbor, as illustrated in Fig.~\ref{fig:structure}. While all three of these interactions $J_\perp$, $J_{\parallel1}$, and $J_{\parallel2}$ are ferromagnetic in Cr$_{1/3}$NbS$_{2}$ and Mn$_{1/3}$NbS$_{2}$, giving rise to the exotic chiral magnetic states, it has been predicted that $J_\perp$ and $J_{\parallel1}$ become antiferromagnetic in Fe$_{1/3}$NbS$_{2}$ \cite{16_mankovsky}. Based on the magnetic structures found for V$_{1/3}$TaS$_{2}$ and V$_{1/3}$NbS$_{2}$ in the current work, a third situation seems to be encountered here: a ferromagnetic $J_\perp$ and a large, antiferromagnetic $J_{\parallel1}$. Within instrument resolution, we have not observed any evidence for magnetic structure modulation caused by a competition between the DM interaction and these exchange couplings. Although our work rules out the possibility of identifying a chiral soliton lattice in the vanadium-intercalated TMDC compounds, we cannot discount the possibility of exotic antiferromagnetic analogs as predicted for skyrmion materials \cite{legrand_2019_roomtemperature,okubo_2012_multipleqstates,gbel_2017_antiferromagnetic}. Additional theoretical studies and neutron scattering experiments can further elucidate the relationship between chiral spin texture stability and the magnetic Hamiltonians for this family of materials.

\section{Conclusions}
In conclusion, our magnetization and x-ray data indicates that the 32K and 50K magnetic transitions observed in V$_{1/3}$TaS$_2$ and V$_{1/3}$NbS$_2$ respectively are correlated with the non-centrosymmetric structure, allowing for a DM interaction. The $T_{c}'$~$>$~100K magnetic transition sometimes observed in these materials is due to intercalant disorder, correlated in scattering probes with a decrease in intensity of the (1,0,1) structural Bragg peak. Our neutron diffraction data shows that the zero-field low temperature magnetic structure is best described by an antiferromagnetic stacking of ferromagnetic planes, in direct contrast to previous reports. The weak ferromagnetic behavior observed in bulk measurements is likely due to a small canting of vanadium spins along the out-of-plane direction. Our work indicates that the exchange interactions in these vanadium systems are significantly modified as compared to well-studied Cr$_{1/3}$NbS$_{2}$ and therefore chiral soliton physics is not expected. We also anticipate that our general discussion and methodology presented here regarding identification of the non-centrosymmetric crystal structures will expedite progress in characterizing this family of interesting monoaxial chiral magnets.

\begin{acknowledgments}
This work was sponsored by the National Science Foundation, under Grant No. DMR-1455264-CAR (G.J.M. and K.L.). D.M. acknowledges support from the Gordon and Betty Moore Foundation’s EPiQS Initiative, Grant GBMF9069. Synthesis, powder X-ray, and magnetization measurements were carried out in part in the Materials Research Laboratory Central Research Facilities, University of Illinois. The single crystal X-ray diffraction was done in the George L. Clark X-Ray Facility in Noyes Laboratory, University of Illinois. This research used resources at the High Flux Isotope Reactor, a DOE Office of Science User Facility operated by the Oak Ridge National laboratory.
\end{acknowledgments}

\appendix*
\renewcommand{\thefigure}{A\arabic{figure}}
\setcounter{figure}{0}
\section{the correlation between the (1,0,1) Bragg peak and V site disorder}
\begin{figure*}[htbp]
\includegraphics[width=1\textwidth]{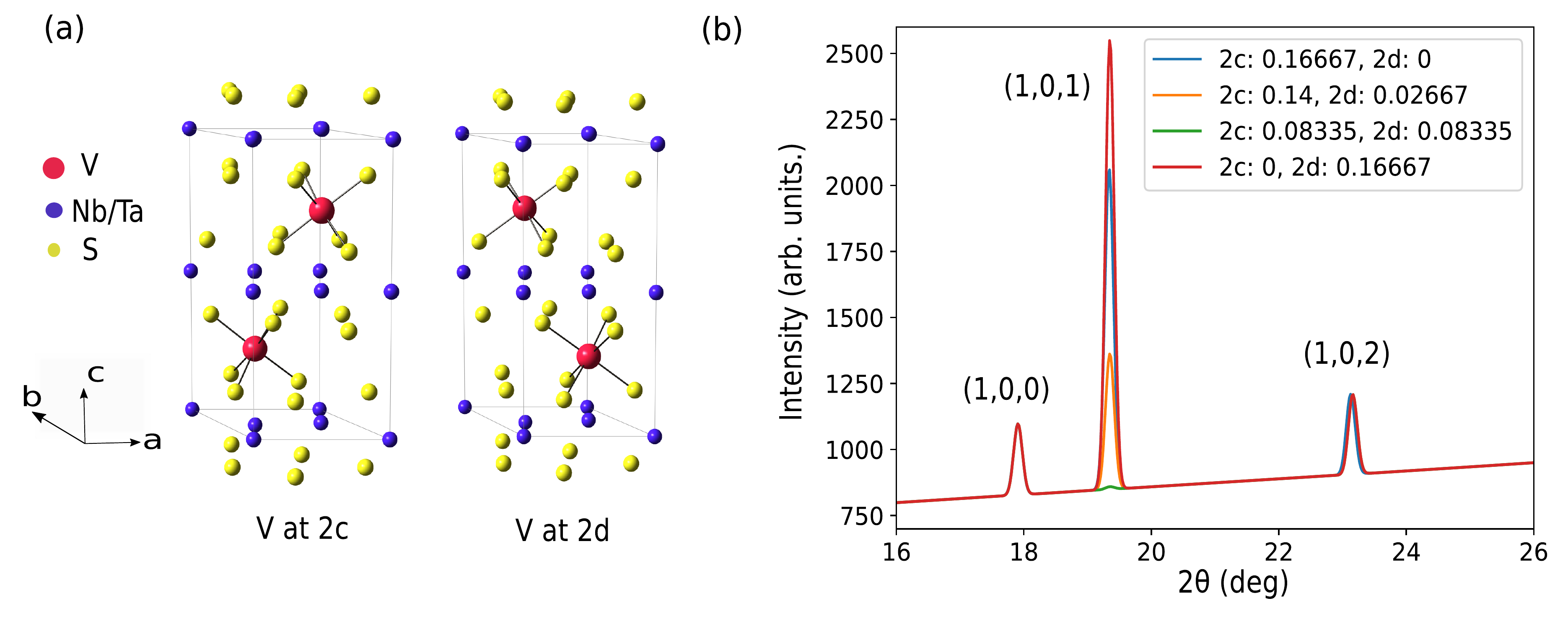}
 \caption{\label{fig:supplement} (a) Crystal structures of V$_{1/3}$TaS$_{2}$ and V$_{1/3}$NbS$_{2}$ with the space group {\it P6$_3$22}. Vanadium can intercalate into the host TMDC at either the 2c or 2d Wyckoff position to break inversion symmetry. (b) Simulated powder x-ray diffraction patterns with intercalated vanadium at different positions. }
 \label{figA1}
\end{figure*}

As discussed in the article, the (1,0,1) Bragg peak can be regarded as a signature of the inversion symmetry breaking. In this Appendix, we use simulated powder x-ray diffraction patterns to show this. When the vanadium atoms are intercalated into the 2H-polytype of NbS$_2$ or TaS$_2$, the intercalants usually occupy the 2c or 2d Wyckoff positions as shown in Fig.~\ref{fig:supplement}(a). The resulting crystal structure then breaks inversion symmetry and allows for the Dzyaloshinskii-Moriya (DM) interaction. Significant site disorder is also possible in some cases, ensuring that the intercalants occupy both the 2c and 2d Wyckoff positions. As a result of this site disorder, the DM interaction is suppressed.

To model a system with disordered intercalants using the FullProf software suite, we can allow the vanadium ions to occupy both the 2c and 2d Wyckoff sites of the space group {\it P6$_3$22} with the total occupancy set to 1/6. We see that the relative intensity of the (1,0,1) Bragg peak is the strongest when the intercalants occupy either the 2c or 2d Wyckoff position only, as shown in Fig.~\ref{fig:supplement}(b). Once we allow simultaneous occupancy of both the 2c and 2d Wyckoff positions, a decrease of the (1,0,1) intensity is observed. Therefore, the (1,0,1) nuclear Bragg peak can be regarded as a signature for the non-centrosymmetric space group P6$_{3}$22 with intercalants at 2c Wyckoff sites (or equivalently, at 2d Wyckoff sites). In the extreme case where vanadium ions occupy the 2c and 2d Wyckoff sites evenly, the (1,0,1) Bragg peak vanishes, as shown in Fig.~\ref{fig:supplement}(b). The correlation between intercalant site disorder and the intensity of the (1,0,1) Bragg peak is invaluable for quickly assessing the likelihood that samples have crystallized in the desirable non-centrosymmetric space group.

\bibliography{VTS_VNS_paper_25May2020}

\end{document}